\documentclass[10pt, pre, aps, twocolumn,showpacs,nofootinbib,longbibliography, dvipdfmx]{revtex4-1}
\usepackage{url,ulem}
\usepackage{amsmath,amssymb}
\usepackage{graphicx}

\newcommand{\R}{\mathbb{R}}

\newcommand{\mK}{\boldsymbol{\rm K}}

\newcommand{\mB}{\boldsymbol{\rm B}}

\newcommand{\mO}{\boldsymbol{\rm O}}
\newcommand{\mP}{\boldsymbol{\rm P}}

\newcommand{\mLambda}{\boldsymbol{\rm \Lambda}}

\newcommand{\mE}{\boldsymbol{\rm E}}

\newcommand{\br}{\boldsymbol{r}}

\newcommand{\bp}{\boldsymbol{p}}

\newcommand{\bv}{\boldsymbol{v}}

\newcommand{\by}{\boldsymbol{y}}

\newcommand{\bzero}{\boldsymbol{0}}
\newcommand{\bbeta}{\boldsymbol{\eta}}

\newcommand{\norm}[1]{||#1||}
\newcommand{\dfracd}[2]{\dfrac{{\rm d} #1}{{\rm d} #2}}
\newcommand{\dfracp}[2]{\dfrac{\partial #1}{\partial #2}}
\newcommand{\dfracpp}[3]{\dfrac{\partial^{2} #1}{\partial #2\partial #3}}
\newcommand{\ave}[1]{\left\langle #1 \right\rangle}
\newcommand{\Tr}{{\rm Tr}}
\newcommand{\wt}{\widetilde}

\newcommand{\epsilonLJ}{\epsilon_{\rm LJ}}
\newcommand{\ls}{l^{\ddag}}
\newcommand{\bys}{\by^{\ddag}}
\newcommand{\ULJ}{U_{\rm LJ}}

\begin{document}

\title{Stabilization of a transition state by excited vibration
  and impact on the reaction rate in the three-body Lennard-Jones system}
\author{Yoshiyuki Y. Yamaguchi}
\email{yyama@amp.i.kyoto-u.ac.jp}
\affiliation{Graduate School of Informatics, Kyoto University, Kyoto 606-8501, Japan}

\begin{abstract}
  The three-body Lennard-Jones system on the plane has a transition state,
  which is the straight conformation located at a saddle point of the potential energy landscape.
  We show that the transition state can be dynamically stabilized
  by excited vibration of particle distances.
  The stabilization mechanism is explained theoretically,
  and is verified by performing molecular dynamics simulations.
  We also examine whether the dynamical stabilization gives an impact on the reaction rate
  between the two isomers of equilateral triangle conformations  
  by comparing with the transition state theory.
\end{abstract}
\maketitle

\section{Introduction}
\label{sec:Introduction}

The state is said to be unstable if the potential energy landscape takes
a saddle or a local maximum at the corresponding state.
A typical example is the inverted pendulum under the gravity.
However, dynamics may change stability,
and the inverted pendulum is stabilized
by applying fast small oscillation of the pivot in the vertical direction.
This pendulum is known as the Kapitza pendulum
\cite{stephenson-08,kapitza-51,butikov-01,bukov-dalessio-polkovnikov-15}.
The essence of this dynamical stabilization is existence of multiple scales in time and space:
The averaged fast small motion gives an additional term to the potential energy landscape,
and the effective potential for slow motion may have a local minimum at the inverted position.
Interestingly, the vertical motion of the pivot contributes to the stabilization
in the orthogonal rotational direction.

The Kapitza pendulum is a nonautonomous system,
but dynamical stabilization has been generalized to autonomous Hamiltonian systems
\cite{yamaguchi-etal-22,yamaguchi-23}.
Consider a chainlike model consisting of three beads and two springs,
where two nearby beads are connected by a spring \cite{rouse-53}.
Comparing with the Kapitza pendulum, the two springs play the role of the oscillating pivot,
and the angle between the two springs corresponds to the pendulum angle.
Focusing on the angle motion,
we can reduce the full system to a one-degree-of-freedom system,
and the reduced part provides an additional potential,
which may change the stability ruled by the bare potential energy landscape.
This phenomenon is named by the dynamically induced conformation (DIC).
It might worth mentioning that vibration in the radial direction
stabilizes the orthogonal rotational direction as the Kapitza pendulum.

DIC is firstly observed in numerical simulations of chainlike bead-spring systems
\cite{yanagita-konishi-jp},
and is explained theoretically in three-body system \cite{yamaguchi-etal-22}
and later in $N$-body systems \cite{yamaguchi-23}
by using the multiple-scale analysis \cite{bender-orszag-99} and the averaging method
\cite{krylov-bogoliubov-34,krylov-bogoliubov-47,guckenheimer-holmes-83}.
A remarkable difference of DIC from the Kapitza pendulum is
that the stabilization depends on excited modes of the vibration:
The mode having the lowest eigenfrequency contributes to the stabilization,
and contribution to the destabilization emerges by exciting
higher eigenfrequency modes \cite{yamaguchi-etal-22,yamaguchi-23}.

The previous studies on DIC have two restrictions:
(i) the system should be chainlike and
(ii) the bending potential which determines conformation
is separately introduced in addition to the spring potential.
Due to these restrictions, a system is out of range
which has pairwise potentials but has no explicit bending potential,
although such a system is important to study.
For instance, the Lennard-Jones potential
\cite{lennard-jones-31,fischer-wendland-23,lenhard-stephan-hasse-24}
is frequently used to model the intermolecular interaction.

In this article we consider the three-body Lennard-Jones system on the plane,
which is not a chainlike model but a ringlike model.
The system has two minima, each of which corresponds to an equilateral triangle.
Let the beads be named by A, B, and C.
One triangle is ABC, and the other is ACB.
Due to two-dimensionality, between the two minima,
there exists a transition state at which conformation is straight.
The transition state is located at a saddle point of the potential energy landscape,
but stabilization of the transition state is numerically observed
and phenomenologically analyzed \cite{shimizu-pc}.

One of the aims of this article is to give a systematic explanation
to the numerically observed stabilization by resulting in DIC.
The other aim is to examine if DIC modifies the reaction rate
which is predicted by the transition state theory (TST)
\cite{marcus-rice-51,marcus-52,rosenstock-etal-52,magee-52,giddings-eyring-54,wieder-marcus-62,laidler-king-83,truhlar-garrett-klippenstein-96}.

The three-body Lennard-Jones system on the plane is a simple model of isomerization,
and it is a first step to deeply understand dynamical effects of isomerization in complex molecules.
For example, excited mode dependence of isomer population has been examined experimentally
\cite{dian-longarte-zweier-02,dian-longarte-winter-zwier-04}.
Further, related dynamical effects have been observed in atomic clusters
by considering the internal coordinates
\cite{yanao-takatsuka-03,yanao-takatsuka-04,yanao-takatsuka-05,yanao-etal-07}.

This paper is organized as follows.
The three-body Lennard-Jones system is introduced
and the potential energy landscape is analyzed in Sec.~\ref{sec:Model}.
We show in Sec.~\ref{sec:multiscales} that the essence of DIC,
separation of temporal and spatial scales,
is realized intrinsically in the three-body Lennard-Jones system.
Dynamically induced stability is studied in Sec.~\ref{sec:DIC},
and is examined in numerical simulations in Sec.~\ref{sec:numerics}.
The reaction rate is reported in Sec.~\ref{sec:transition-rate}.
The last section \ref{sec:summary} is devoted to summary.

\section{Model}
\label{sec:Model}

\begin{figure}
  \centering
  \includegraphics[width=5.0cm]{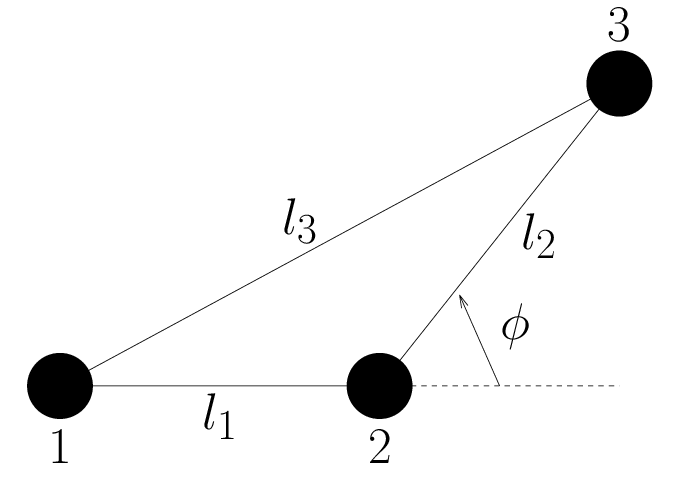}
  \caption{Schematic figure of the three-body system
    with the definitions of $l_{1},l_{2},l_{3}$, and $\phi$.}
  \label{fig:System}
\end{figure}

We consider the three-body Lennard-Jones system on $\R^{2}$.
The system is described by the Lagrangian
\begin{equation}
  L(\br,\dot{\br}) = \dfrac{1}{2} \sum_{i=1}^{3} m\norm{\dot{\br_{i}}}^{2} - V(\br),
\end{equation}
where $\br_{i}\in\R^{2}$ is the position of the $i$th mass,
$\dot{\br}_{i}=d\br_{i}/dt$, $t$ is the time,
and $\br=(\br_{1},\br_{2},\br_{3})$.
We assume that the three masses are identical.
The total potential energy $V$ consists of pairwise interactions as
\begin{equation}
  \label{eq:V}
  V(\br) = \ULJ(l_{1}) + \ULJ(l_{2}) + \ULJ(l_{3}),
\end{equation}
where
\begin{equation}
  l_{1} = \norm{\br_{1}-\br_{2}},
  \quad
  l_{2} = \norm{\br_{2}-\br_{3}},
  \quad
  l_{3} = \norm{\br_{3}-\br_{1}}.
\end{equation}
A schematic figure of the system is shown in Fig.~\ref{fig:System}.

\begin{figure}
  \centering
    \includegraphics[width=8.5cm]{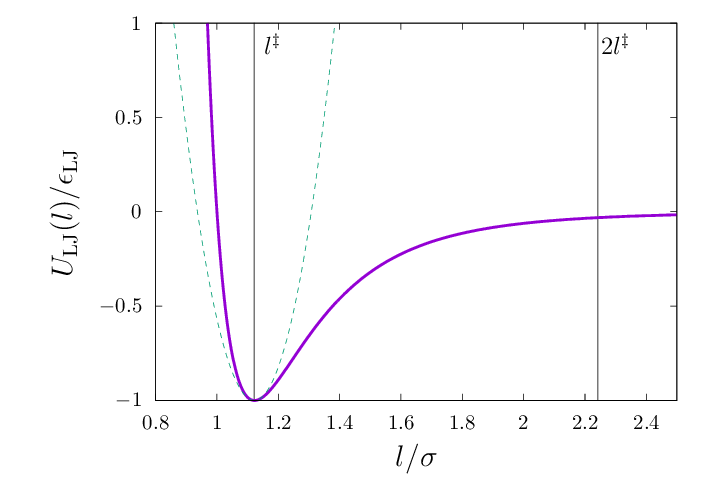}
  \caption{Lennard-Jones potential (purple solid line).
    The lengths at a straight conformation are $l_{1}=l_{2}=\ls$ and $l_{3}=2\ls$,
    which are marked by the vertical black lines.
    The green broken curve represents the harmonic approximation around the bottom point.}
  \label{fig:LennardJones}
\end{figure}

The pairwise potential $\ULJ$ is the Lennard-Jones potential
\begin{equation}
  \ULJ(l) = 4 \epsilonLJ \left[
    \left( \dfrac{\sigma}{l} \right)^{12} - \left( \dfrac{\sigma}{l} \right)^{6} 
  \right],
\end{equation}
where $\sigma$ and $\epsilonLJ$ are real positive parameters.
The equations of motion are
\begin{equation}
  \dfracd{{}^{2}\br_{i}}{t^{2}} = - \dfrac{4\epsilonLJ}{m}
  \sum_{j=1}^{3} \dfracp{}{\br_{i}} \left[ \left( \dfrac{\sigma}{l_{j}} \right)^{12}
    - \left( \dfrac{\sigma}{l_{j}} \right)^{6} \right],
  ~(i=1,2,3)
  \label{eq:EOM}
\end{equation}
and the values of $\sigma$ and $\epsilon_{\rm LJ}/m$ are not essential,
because we can set them as unity by rescaling the space $\br_{i}$ and the time $t$.
Landscape of the Lennard-Jones potential is illustrated in Fig.~\ref{fig:LennardJones}.
It has the minimum value $\ULJ(l_{\rm min})=-\epsilonLJ$ at
\begin{equation}
  l_{\rm min} = a_{\rm min} \sigma,
  \quad
  a_{\rm min}=2^{1/6}=1.122\cdots.
\end{equation}
% $l_{\rm min}=a_{\rm min}\sigma$ with $a_{\rm min}=2^{1/6}=1.122...$.

\subsection{Internal coordinates}
The system has the translational symmetry and the rotatinal symmetry.
The associated integrals are the total momentum vector and the total angular momentum.
We set them as zero.

We reduce the translational symmetry 
by introducing the internal coordinates
$\by=(y_{1},y_{2},y_{3},y_{4})=(l_{1},l_{2},\phi,\phi_{\rm L})$.
The angle $\phi$ is the bending angle between $\br_{2}-\br_{1}$ and $\br_{3}-\br_{2}$.
For instance, $\phi=0$ represents a straight conformation (see Fig.~\ref{fig:System}).
The last angle variable $\phi_{\rm L}$ is associated with the total angular momentum,
which is another integral of the system.

The length $l_{3}$ is written as a function of $(l_{1},l_{2},\phi)$ as
\begin{equation}
  \label{eq:l3-2}
  l_{3}^{2} = l_{1}^{2} + l_{2}^{2} + 2l_{1}l_{2}\cos\phi.
\end{equation}
The total potential energy $V$ is read as
\begin{equation}
  V(\by)
  = \ULJ(l_{1}) + \ULJ(l_{2}) + \ULJ\left(\sqrt{l_{1}^{2}+l_{2}^{2}+2l_{1}l_{2}\cos\phi}\right).
\end{equation}
The Lagrangian in the internal coordinates is
\begin{equation}
  \label{eq:Lagrangian}
  L(\by,\dot{\by}) = \dfrac{1}{2} \sum_{\alpha,\beta=1}^{4}
  B^{\alpha\beta}(\by) \dot{y}_{\alpha} \dot{y}_{\beta} - V(\by),
\end{equation}
where $B^{\alpha\beta}(\by)$ is the $(\alpha,\beta)$ element of the matrix $\mB(\by)$.
See Appendix \ref{sec:matrixB} for the explicit expression of $\mB(\by)$.
The variable $\phi_{\rm L}$ is not crucial since it is a cyclic coordinate
associated with the rotational symmetry.
Hereafter we omit the angle $\phi_{\rm L}$, 
and use the same symbol $\by$ for $\by=(y_{1},y_{2},y_{3})=(l_{1},l_{2},\phi)$.

\subsection{Stationary points of $V$}

The potential $V$ has two minima and three saddle points.
The two minima are located at
$(l_{1},l_{2},\phi)=(l_{\rm min},l_{\rm min},2\pi/3)$
and $(l_{\rm min},l_{\rm min},-2\pi/3)$,
which correspond to the equilateral triangle and the inverse equilateral triangle conformations.
Between the two minima, there exists a saddle at $\bys=(\ls,\ls,0)$,
which represents a straight conformation and which we focus on.
We omit two other saddles (straight conformations)
since they are obtained by changing the order of masses.
The length $\ls$ satisfies the stationarity condition
\begin{equation}
  \dfracp{V}{l_{i}}(\bys) = \ULJ'(\ls) + \ULJ'(2\ls) = 0,
  \quad
  i=1,2
  \label{eq:straight-stationarity-condition}
\end{equation}
since $l_{3}=2\ls$ at the saddle (See Fig.~\ref{fig:LennardJones}). The length $\ls$ is
\begin{equation}
  \ls = a\sigma,
  \quad
  a = \dfrac{1}{2} \left[ \dfrac{2(2^{13}+1)}{2^{7}+1} \right]^{1/6}
  = 1.121...
\end{equation}
We note that $\ls$ is smaller than and close to $l_{\rm min}$.
The values of $V(\by)$ at the minima and the saddle are
\begin{equation}
  \dfrac{V(l_{\rm min},l_{\rm min},\pm 2\pi/3)}{\epsilonLJ} = -3,
  \quad
  \dfrac{V(\bys)}{\epsilonLJ} \simeq -2.031124.
\end{equation}

\begin{figure}
  \centering
  \includegraphics[width=8.5cm]{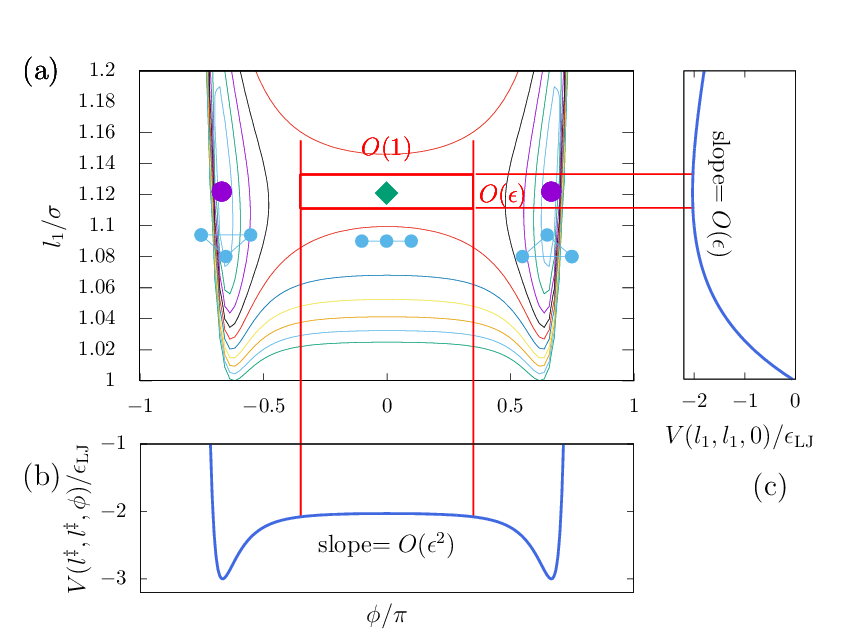}
  \caption{Potential energy landscape of the three-body Lennard-Jones system
    on the section $l_{2}=l_{1}$, i.e. $V(l_{1},l_{1},\phi)$.
    (a) Contours (curves), two minima (purple circles), and a saddle (green diamond).
    Conformations are presented by blue beads and lines.
    The red rectangle is the region in which we are interested.
    (b) Potential $V(\ls,\ls,\phi)/\epsilonLJ$ as a function of $\phi$.
    The slope is of $O(\epsilon^{2})$ since $V(\by)-V(\bys)=O(\epsilon^{2})$
    in the interested region.
    (c) Potential $V(l_{1},l_{1},0)/\epsilonLJ$ as a function of $l_{1}$.
    The slope is of $O(\epsilon)$ in the interested region.
    See Eq.~\eqref{eq:epsilon} for the definition of the small parameter $\epsilon$.
  }
  \label{fig:potential-contour}
\end{figure}

The two minima and the saddle $\bys$, which is a transition state,
are illustrated in Fig.~\ref{fig:potential-contour}
with a contour map of $V(l_{1},l_{2},\phi)$ on the section $l_{2}=l_{1}$.
The straight conformation is unstable for the $\phi$ direction
with respect to the bare potential $V$,
the conformation can be stabilized by fast vibration of $l_{1}$ and $l_{2}$ however:
This is DIC.

\section{Multiscale analysis}
\label{sec:multiscales}

The essential mechanism of DIC is separation of temporal and spatial scales.
In the previous study \cite{yamaguchi-etal-22, yamaguchi-23},
to realize slow bending motion of $\phi$,
the total potential is composed of
the spring potential $V_{\rm spring}(l_{1},l_{2})$
and the bending potential $\epsilon^{2}V_{\rm bend}(\phi)$,
where $\epsilon$ is a small dimensionless parameter $\epsilon~(0<\epsilon\ll 1)$.
We demonstrate that the three-body Lennard-Jones system
induces intrinsically the separation of scales and a bending potential of $O(\epsilon^{2})$
around the transition state $\bys$.

\subsection{Separation of timescales}
We estimate the timescales around the transition state $\bys$
by considering the linearized equations of motion.
The Taylor expansion of the total potential $V$ around $\bys$ gives
\begin{equation}
  \label{eq:nablaV-expansion}
  \nabla V(\by) = H(\bys) (\by-\bys) + \cdots,
\end{equation}
where
\begin{equation}
  \label{eq:Kbys}
  \begin{split}
    H(\bys)
    & = \ULJ''(\ls)
      \begin{pmatrix}
        1 & 0 & 0 \\
        0 & 1 & 0 \\
        0 & 0 & 0 \\
      \end{pmatrix}
      + \ULJ''(2\ls)
      \begin{pmatrix}
        1 & 1 & 0 \\
        1 & 1 & 0 \\
        0 & 0 & 0 \\
      \end{pmatrix}
      \\
    & - \dfrac{\ULJ'(2\ls)}{2\ls} (\ls)^{2}
      \begin{pmatrix}
        0 & 0 & 0 \\
        0 & 0 & 0 \\
        0 & 0 & 1 \\
      \end{pmatrix}
  \end{split}
\end{equation}
is the Hessian of $V$ at $\by=\bys$.
 
The Hessian matrix implies that
the timescale $\tau_{l}$ of $l_{i}$ is determined by the first and the second terms of $H$,
and that the timescale $\tau_{\phi}$ of $\phi$ by the third term.
In the third term we extracted $(\ls)^{2}$ to adjust the physical dimension.
The coefficients are estimated as
\begin{equation}
  \label{eq:Uderivatives}
  \begin{split}
    & \ULJ''(\ls) = \dfrac{24\epsilonLJ}{a^{12}(\ls)^{2}} ( 26-7a^{6})
      \simeq \dfrac{24\epsilonLJ}{a^{12}(\ls)^{2}} 12.1068,
    \\
    & \ULJ''(2\ls) = \dfrac{24\epsilonLJ}{a^{12}(\ls)^{2}} \dfrac{26-7(2a)^{6}}{2^{14}}
      \simeq \dfrac{24\epsilonLJ}{a^{12}(\ls)^{2}} (-0.05268), \\
    & \dfrac{\ULJ'(2\ls)}{2\ls} = - \dfrac{24\epsilonLJ}{a^{12}(\ls)^{2}} \dfrac{2-(2a)^{6}}{2^{14}}
      \simeq \dfrac{24\epsilonLJ}{a^{12}(\ls)^{2}} 0.00763. \\
  \end{split}
\end{equation}
The first term $\ULJ''(\ls)$ dominates the second term $\ULJ''(2\ls)$,
and the timescales are estimated as
\begin{equation}
  \tau_{l} = \sqrt{\dfrac{m}{\ULJ''(\ls)}},
  \qquad
  \tau_{\phi} = \sqrt{\dfrac{m}{\ULJ'(2\ls)/(2\ls)}}.
\end{equation}
Separation of the two timescales is now clear,
and we introduce a small dimensionless parameter $\epsilon$ as
\begin{equation}
  \epsilon = \dfrac{\tau_{l}}{\tau_{\phi}}
  = \sqrt{\dfrac{\ULJ'(2\ls)/(2\ls)}{\ULJ''(\ls)}} \simeq 0.0251.
  \label{eq:epsilon}
\end{equation}

The small parameter $\epsilon$ introduces multiple timescales:
the fast timescale $t_{0}=t$ corresponding to $\tau_{l}$
and the slow timescale $t_{1}=\epsilon t$ corresponding to $\tau_{\phi}$.
We consider that $t_{0}$ and $t_{1}$ are independent variables.
The time derivative is expanded as
\begin{equation}
  \label{eq:expansion-t}
  \dfracd{}{t} = \dfracp{}{t_{0}} + \epsilon \dfracp{}{t_{1}}.
\end{equation}

\subsection{Spatial scale and potential forces}
For the dependent variables, we are interested in a region close to the transition state $\bys$
and, from Fig.~\ref{fig:potential-contour}, we expand them as
\begin{subequations}
  \label{eq:expansion-l-phi}
  \begin{align}
    \label{eq:expansion-l}
    & l_{i}(t_{0},t_{1}) = \ls + \epsilon l_{i}^{(1)}(t_{0},t_{1}), \quad i=1,2 \\
    \label{eq:expansion-phi}
    & \phi(t_{0},t_{1}) = \phi^{(0)}(t_{1}) + \epsilon \phi^{(1)}(t_{0},t_{1}),
  \end{align}
\end{subequations}
which is simply denoted by
\begin{equation}
  \label{eq:expansion-y}
  \by(t_{0},t_{1}) = \by^{(0)}(t_{1}) + \epsilon \by^{(1)}(t_{0},t_{1}).
\end{equation}
Smallness of $|\epsilon l_{i}^{(1)}|$ is realized by selecting suitable initial conditions.

We consider the potential force again, which is expanded as
\begin{equation}
  \label{eq:expansion-gradV}
  \nabla V(\by)
  = \epsilon (\nabla V(\by))^{(1)} + \epsilon^{2} (\nabla V(\by))^{(2)} + \cdots,
\end{equation}
where, for an arbitrary function $A(\by)$,
$(A(\by))^{(n)}$ represents the $O(\epsilon^{n})$ part of $A(\by)$.
From Eqs.~\eqref{eq:nablaV-expansion}, \eqref{eq:Kbys}, and \eqref{eq:epsilon}, we have
\begin{equation}
  \label{eq:expansion-gradV-1}
  (\nabla V(\by))^{(1)} = \ULJ''(\ls)
  \begin{pmatrix}
    l_{1}^{(1)} \\ l_{2}^{(1)} \\ 0 \\ 
  \end{pmatrix}
\end{equation}
and
\begin{equation}
  \label{eq:expansion-gradV-2}
  (\nabla V(\by))^{(2)} = - \ULJ''(\ls) (\ls)^{2}
  \begin{pmatrix}
    \ast \\ \ast \\ \phi \\
  \end{pmatrix}.
\end{equation}
The asterisk parts are not important in the later discussions and we neglect them.
The bending potential depending on $\phi$ appears from $O(\epsilon^{2})$
without any artificial scalings.
The two scales of $\nabla V$ can be understood graphically in Fig.~\ref{fig:potential-contour}
with the fact $V(\by)-V(\bys)=O(\epsilon^{2})$ around $\by^{\ddag}$.

\subsection{Universality}
The expansions \eqref{eq:expansion-t}, \eqref{eq:expansion-l-phi},
and \eqref{eq:expansion-gradV-1} are the same as ones used in the previous analysis
\cite{yamaguchi-etal-22, yamaguchi-23},
whereas the three-body Lennard-Jones system does not satisfy
the two restrictions mentioned in Sec.~\ref{sec:Introduction}.
Therefore, DIC should be reproduced around the transition state $\bys$.

Let us revisit the origin of the above expansions.
The two timescales and $(\partial V/\partial\phi)^{(1)}=0$ are induced
from smallness of the ratio \eqref{eq:epsilon}.
Thus, we may expect emergence of DIC
in other systems apart from the Lennard-Jones potential,
if the pairwise potential $U$ has a steep well [i.e. large $U''(\ls)$]
and a gradual tail [i.e. small $U'(2\ls)/(2\ls)$].
A harmonic potential $U(l)=k(l-\sigma)^{2}/2$ is out of this scope,
since $U''(\ls)=U''(2\ls)=k$ and $U'(2\ls)/(2\ls)=k/4$ are of the same order,
where the stationary length $\ls$ satisfying Eq.~\eqref{eq:straight-stationarity-condition}
is $\ls=2\sigma/3$.

\section{Dynamical stability of the transition state}
\label{sec:DIC}
Along computations performed in \cite{yamaguchi-etal-22,yamaguchi-23},
we can reduce the three-dimensional dynamical system of $(l_{1},l_{2},\phi)$
to the one-dimensional dynamical system of $\phi^{(0)}(t_{1})$,
which appears in $O(\epsilon^{2})$. We sketch the derivation.
See Appendix \ref{sec:EL-eqs} for details.

In $O(\epsilon)$, we have linear equations of motion
for fast motion of $l_{1}^{(1)}$ and $l_{2}^{(1)}$.
The two lengths $l_{1}$ and $l_{2}$ extend and contract simultaneously
in the in-phase mode (mode-I) and alternatively in the antiphase mode (mode-II).
If we modify $l_{1}$ and $l_{2}$ initially from the equilibrium value $\ls$ as
\begin{equation}
  l_{i} = \ls + \delta l_{i}, \quad (i=1,2)
\end{equation}
the initial values of the mode-I energy $E_{\rm I}$ and the mode-II energy $E_{\rm II}$ are
\begin{equation}
  E_{\rm I} = \dfrac{\ULJ''(\ls)}{4} (\delta l_{1}+\delta l_{2})^{2},
  \quad
  E_{\rm II} = \dfrac{\ULJ''(\ls)}{4} (\delta l_{1}-\delta l_{2})^{2}.
  \label{eq:EI-EII}
\end{equation}
We hypothesize that
the ratio between the two normal mode energy $E_{\rm I}$
and $E_{\rm II}$ is constant in time \cite{yamaguchi-etal-22,yamaguchi-23},
and introduce $\nu_{\rm I}:\nu_{\rm II}=E_{\rm I}:E_{\rm II}$ with $\nu_{\rm I}+\nu_{\rm II}=1$.

The equation of motion for $\phi^{(0)}(t_{1})$ is obtained in $O(\epsilon^{2})$,
which includes the fast motion of $l_{1}^{(1)}(t_{0},t_{1})$ and $l_{2}^{(1)}(t_{0},t_{1})$.
Performing the averaging over the fast timescale $t_{0}$,
we can eliminate the initial phases of the two normal modes,
but cannot eliminate the energy ratios, $\nu_{\rm I}$ and $\nu_{\rm II}$.
This is the origin of the excited mode dependence in DIC.

The above-mentioned averaging induces an extra force to $\phi^{(0)}(t_{1})$
in addition to the potential gradient.
The effective force $F_{\rm eff}$ of $O(\epsilon^{0})$
to the slow motion of $\phi^{(0)}(t_{1})$ is expressed as
\begin{equation}
  \epsilon^{2} F_{\rm eff}(\phi^{(0)})
  = - \dfracp{V}{\phi}(\ls,\ls,\phi^{(0)}) - \dfrac{E_{\rm normal}}{2} T(\phi^{(0)}).
  \label{eq:Feff}
\end{equation}
We remark that the potential force $\partial V/\partial\phi$ in Eq.~\eqref{eq:Feff}
is of $O(\epsilon^{2})$ from Eqs.~\eqref{eq:expansion-gradV-1} and \eqref{eq:expansion-gradV-2},
and $E_{\rm normal}=E_{\rm I}+E_{\rm II}$ is also of $O(\epsilon^{2})$ from \eqref{eq:EI-EII}.
The function $T(\phi^{(0)})$ is defined by
\begin{equation}
  T(\phi^{(0)}) = \sin\phi^{(0)} \left(
    \dfrac{\nu_{{\rm I}}}{2-\cos\phi^{(0)}}
    - \dfrac{\nu_{{\rm II}}}{2+\cos\phi^{(0)}} \right).
\end{equation}

\begin{figure}
  \centering
  \includegraphics[width=8cm]{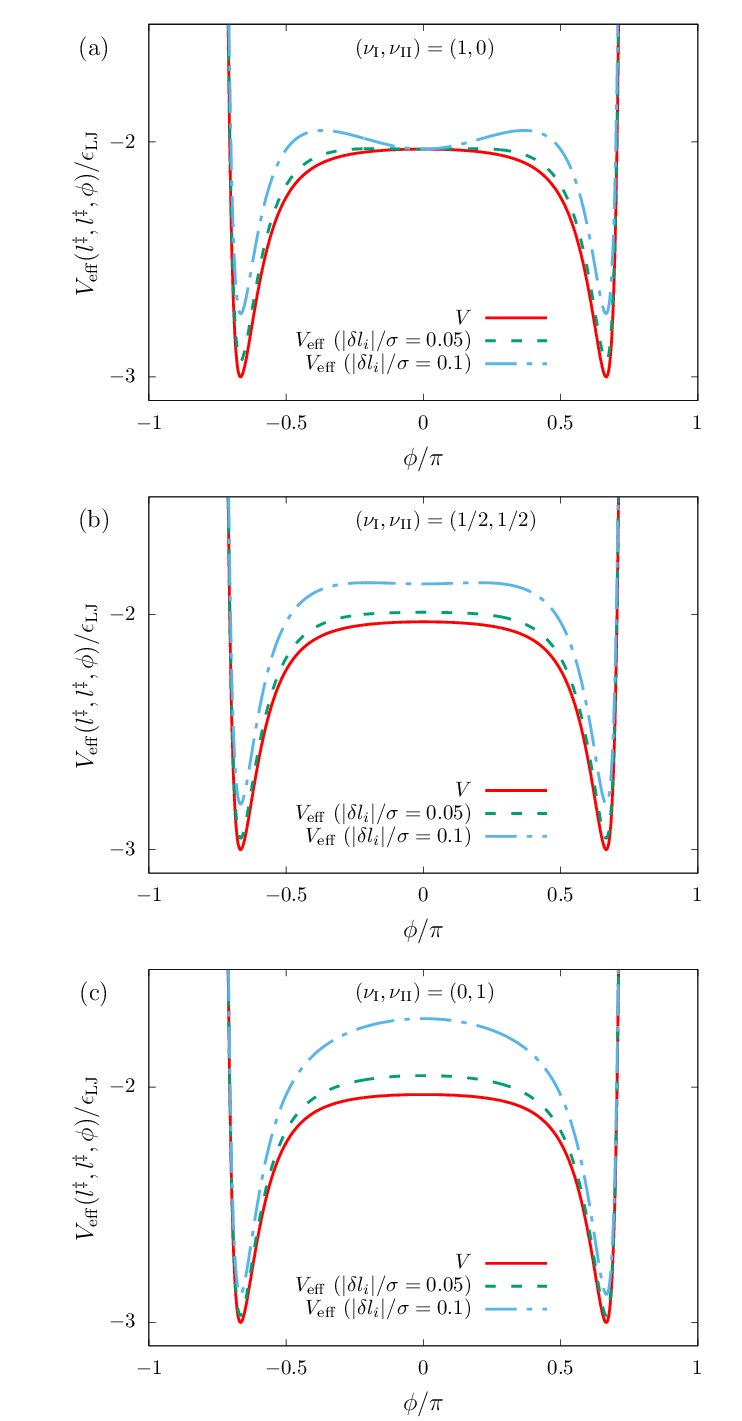}
  \caption{Effective potential $V_{\rm eff}(\phi)$.
    Bare potential $V$ (red solid), effective potential $V_{\rm eff}$
    with $|\delta l_{1}|/\sigma=|\delta l_{2}|/\sigma=0.05$ (green broken)
    and $0.1$ (blue dot-dashed).
    (a) In-phase mode $(\nu_{\rm I},\nu_{\rm II})=(1,0)$.
    (b) Equally mixed mode $(\nu_{\rm I},\nu_{\rm II})=(1/2,1/2)$.
    (c) Antiphase mode $(\nu_{\rm I},\nu_{\rm II})=(0,1)$.
  }
  \label{fig:Veff}
\end{figure}

Integrating the effective force \eqref{eq:Feff},
we have the effective potential $V_{\rm eff}(\phi)$ for the angle $\phi$ as
\begin{equation}
  \begin{split}
    & V_{\rm eff}(\phi)
      = V(\ls,\ls,\phi)\\
    & + \dfrac{E_{\rm normal}}{2} \left[
      \nu_{\rm I} \log(2-\cos\phi)
      + \nu_{\rm II} \log(2+\cos\phi)
      \right].
  \end{split}
  \label{eq:Veff}
\end{equation}
Examples of the effective potential $V_{\rm eff}$ are shown in Fig.~\ref{fig:Veff}.
The in-phase mode stabilizes the straight conformation $\phi=0$,
while the antiphase mode enhances instability.
This mode dependence is the same as the chainlike model \cite{yamaguchi-etal-22,yamaguchi-23}.
The bare potential $V$ is of $O(\epsilon^{0})$ and $E_{\rm normal}$ is of $O(\epsilon^{2})$,
the additional potential function is visible however for $|\delta_{i}|/\sigma\simeq 0.05$,
since $\ULJ''(\ls)/4$ appearing in \eqref{eq:EI-EII} gives a factor
\begin{equation}
  \dfrac{\ULJ''(\ls)}{4} \simeq \dfrac{\epsilonLJ}{\sigma^{2}} 14.6737
\end{equation}
and $E_{\rm normal}/\epsilonLJ$ is around $0.15$.

We remark an important modification from the previous work \cite{yamaguchi-etal-22}.
In the previous work, the normal mode energy $E_{\rm normal}$ is eliminated
in the effective potential $V_{\rm eff}(\phi)$ by using the total energy conservation,
which includes energy of $\phi^{(0)}$ motion.
The total energy conservation provides a global effective potential
under the hypothesis that the ratio $\nu_{\rm I}:\nu_{\rm II}$ is constant of time globally.
In other words, the obtained effective potential is valid for any value of $\phi$.
The constant ratio hypothesis is not bad in a chainlike model,
but it is not good in the three-body Lennard-Jones system if $\phi$ is far from $0$,
because thermalization occurs easily
(see Sec.~\ref{sec:hypothesis} for a numerical test
and Sec.~\ref{sec:transition-rate} for a success of a statistical theory).
Therefore, we concentrate on a neighborhood of $\phi=0$
in which the hypothesis should be valid,
and the effective potential \eqref{eq:Veff} is valid only around $\phi=0$.

Stability of the transition state $\bys=(\ls,\ls,0)$
is determined by the signature of
\begin{equation}
  V_{\rm eff}''(0)
  = \dfracp{{}^{2}V}{\phi^{2}}(\bys)
  + \dfrac{3\nu_{\rm I}-\nu_{\rm II}}{6} E_{\rm normal}.
  \label{eq:ddVeff-at0}
\end{equation}
It is clear that the in-phase mode (mode-I) contributes to stabilization,
and that the antiphase mode (mode-II) to destabilization.
Using Eqs.~\eqref{eq:V}, \eqref{eq:l3-2}, and \eqref{eq:EI-EII},
we modify $V_{\rm eff}''(0)$ as
\begin{equation}
  V_{\rm eff}''(0) = \dfrac{\sigma^{2}}{12} \ULJ''(\ls) S
\end{equation}
where
\begin{equation}
  S = 
  - 12 \epsilon^{2} a^{2}
  + \dfrac{\delta l_{1}^{2} + \delta l_{2}^{2} + 4 \delta l_{1} \delta l_{2}}{\sigma^{2}}
  \label{eq:stability-condition}
\end{equation}
plays the role of the stability index:
$S>0$ ($S<0$) means stable (unstable) since $\ULJ''(\ls)>0$.
The constant term is
\begin{equation}
  12 \epsilon^{2} a^{2} \simeq 0.0095.
\end{equation}

\section{Molecular dynamics simulations}
\label{sec:numerics}

We examine the theoretical stability $S>0$ [see Eq.~\eqref{eq:stability-condition}]
by comparing it with molecular dynamics (MD) simulations.
We set $m=1$, $\sigma=1$, and $\epsilonLJ=1$ without loss of generality
by rescaling the time $t$ and space $\br_{i}$ in the equations of motion \eqref{eq:EOM},
although we keep the symbols to clarify the source.
MD simulations are performed by using the $4$th order symplectic integrator
\cite{yoshida-90} with the timestep $\Delta t=10^{-3}$.

\subsection{Initial condition}
\label{sec:initial-condition}
The initial velocities of particles are zero, $\dot{\br}_{i}=\bzero~(i=1,2,3)$,
and the initial momenta are zero accordingly.
The initial positions are shifted from the saddle point as
\begin{equation}
  \begin{pmatrix}
    x_{1}(0) \\ x_{2}(0) \\ x_{3}(0)
  \end{pmatrix}
  =
  \begin{pmatrix}
    -\ls+\delta x_{1} \\
    -\delta x_{1}-\delta x_{3} \\
    \ls+\delta x_{3}
  \end{pmatrix},
  \quad
  \begin{pmatrix}
    y_{1}(0) \\ y_{2}(0) \\ y_{3}(0)
  \end{pmatrix}
  =
  \begin{pmatrix}
    -\delta y/2 \\
    \delta y \\
    -\delta y/2 \\
  \end{pmatrix}.
\end{equation}
The displacements $\delta x_{1}$ and $\delta x_{3}$ are transformed to $\delta l_{i}$ as
\begin{equation}
  \delta l_{1} = -2\delta x_{1} - \delta x_{3},
  \qquad
  \delta l_{2} = \delta x_{1} + 2\delta x_{3},
\end{equation}
and we vary $\delta l_{1}/\sigma$ and $\delta l_{2}/\sigma$.
$\delta y=10^{-3}$ is given as a perturbation to modify the value of $\phi$ from zero.
If $\delta l_{1}=\delta l_{2}=\delta y=0$, the state is exactly at the transition state
and no temporal evolution occurs.

\subsection{Temporal evolution}
\begin{figure}
  \centering
    \includegraphics[width=8.5cm]{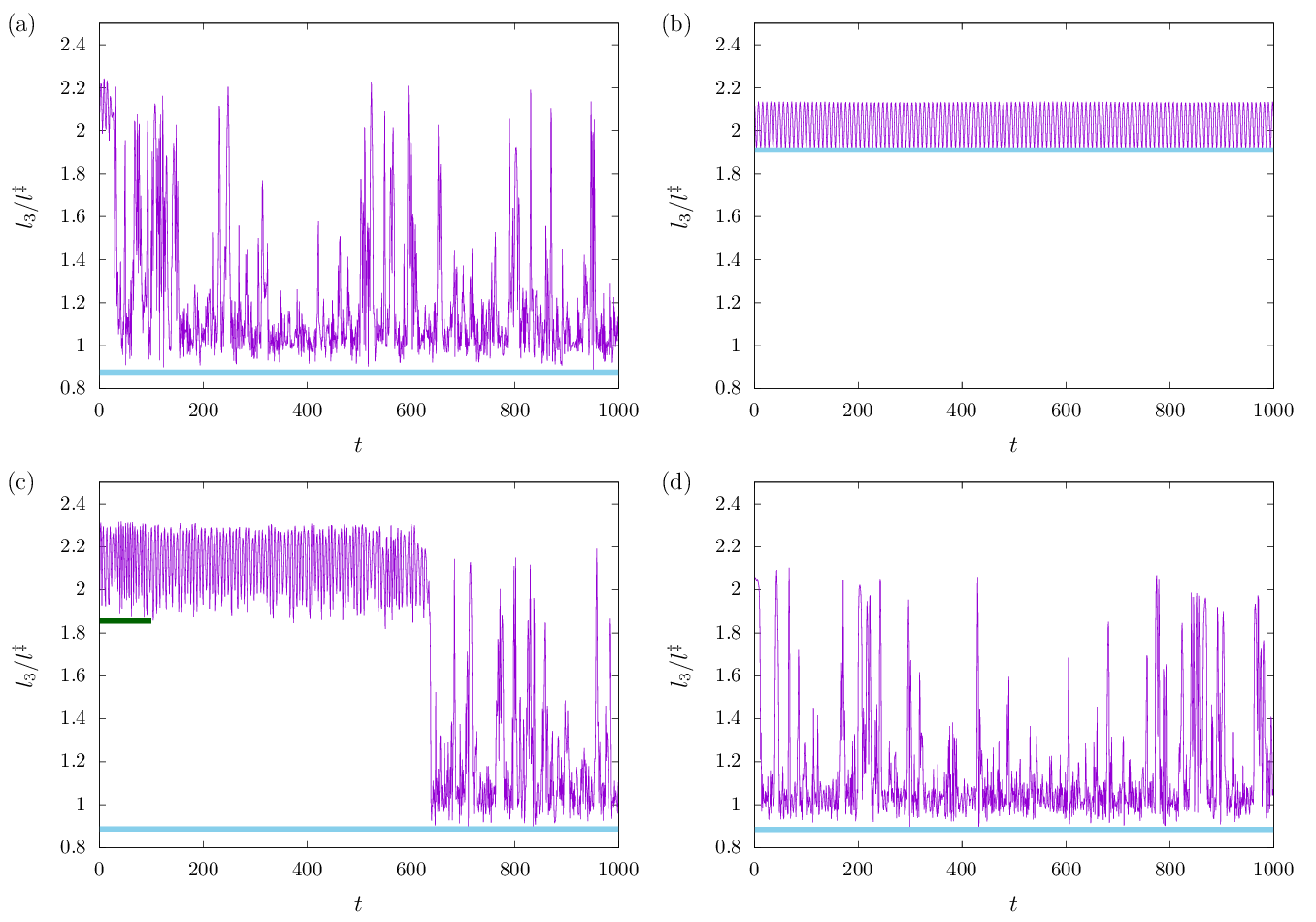}
  \caption{Temporal evolution of $l_{3}/\ls$.
    $(\delta l_{1}/\sigma, \delta l_{2}/\sigma)$ is
    (a) $(-0.1, 0.05)$, (b) $(0.1, 0.05)$, (c) $(-0.1,-0.05)$, and (d) $(0.1,-0.05)$.
    The blue horizontal line represents $\wt{l}_{3,{\rm min}}(10^{3})$,
    and the green short horizontal line represents $\wt{l}_{3,{\rm min}}(10^{2})$ in (c)
    [see Eq.~\eqref{eq:l3min} for $\wt{l}_{3,{\rm min}}(T)$].
    The theoretical stability index $S$ is $S=-0.017$ for (a) and (d),
    and $S=0.023$ for (b) and (c).
  }
  \label{fig:TemporalEvolution}
\end{figure}

First, we observe temporal evolution of $l_{3}/\ls$ shown in Fig.~\ref{fig:TemporalEvolution}.
The quantity $l_{3}/\ls$ is close to $2$ if the system stays around the straight conformation
(a transition state),
and is close to $1$ if the system is around an equilateral triangle conformation
(a minimum).
In other words, $l_{3}/\ls\simeq 2$ implies stabilization of the transition state,
and $l_{3}/\ls\simeq 1$ implies nonstabilization.

The initial excitation of the normal modes is controlled by the signs of $\delta l_{i}$:
$E_{\rm I}>E_{\rm II}$ if the signs are the same,
and $E_{\rm I}<E_{\rm II}$ if the signs are opposite.
In Figs.~\ref{fig:TemporalEvolution}(a) and (d),
the antiphase mode (mode-II) is dominant as $E_{\rm I}/E_{\rm II}=1/9$,
and the straight conformation is not stabilized.
However, in Figs.~\ref{fig:TemporalEvolution}(b) and (c),
the in-phase mode (mode-I) is dominant as $E_{\rm I}/E_{\rm II}=9$,
and the straight conformation is stabilized up to $t\simeq 600$ at least.
Temporal evolution of $l_{3}/\ls$ is consistent with the effective potential
shown in Fig.~\ref{fig:Veff}.

Asymmetry between Figs.~\ref{fig:TemporalEvolution}(b) and (c) is explained as follows.
The Lennard-Jones potential has a steeper wall in the contracting direction $\delta l_{i}<0$
than the extending direction $\delta l_{i}>0$ as shown in Fig.~\ref{fig:LennardJones}.
Nonlinear effects are hence stronger in Fig.~\ref{fig:TemporalEvolution}(c)
than in Fig.~\ref{fig:TemporalEvolution}(b),
while the theory cannot capture them
since it approximates $E_{\rm normal}$ by a harmonic potential.
See also the green broken curve in Fig.~\ref{fig:LennardJones}.

\subsection{Stability diagram}
\begin{figure}
  \centering
    \includegraphics[width=8.5cm]{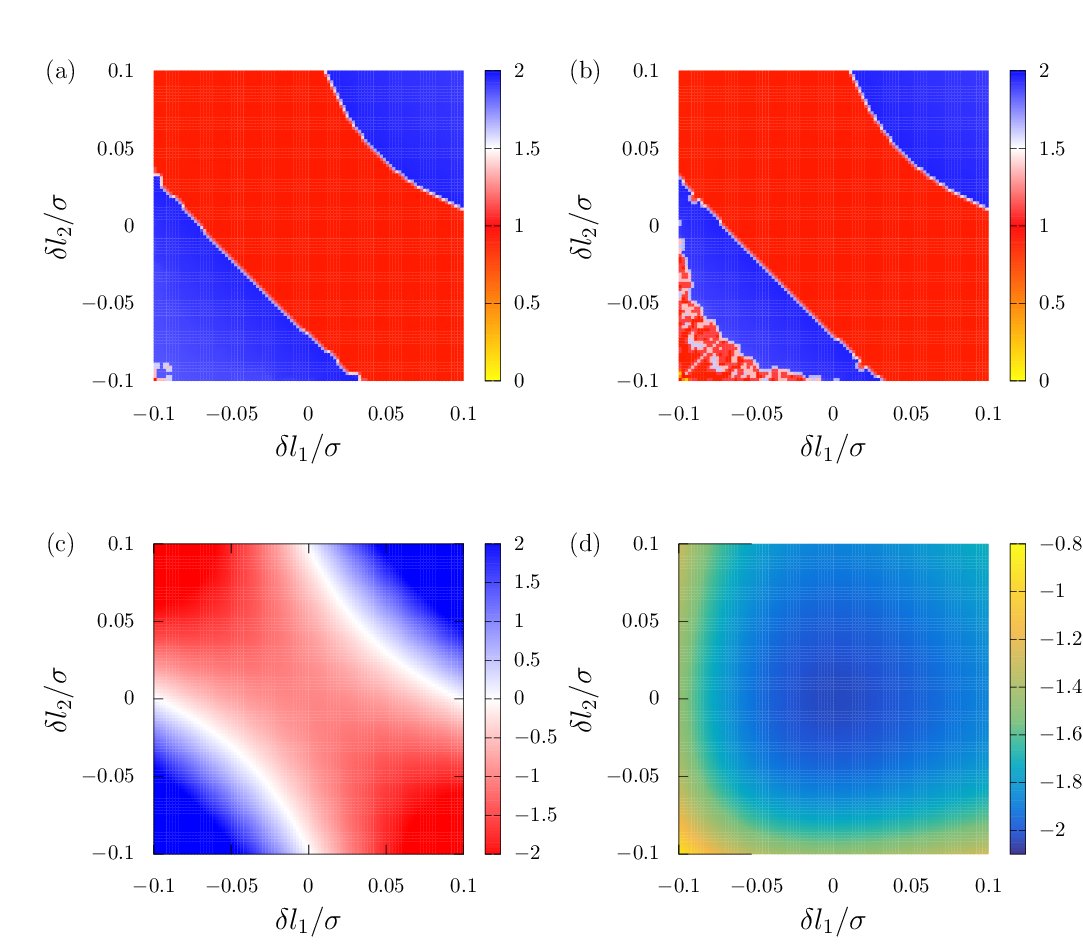}
  \caption{Stability of the transition state.
    (a) $\widetilde{l}_{3,{\rm min}}(T=10^{2})$.
    (b) $\widetilde{l}_{3,{\rm min}}(T=10^{3})$.
    Note $\widetilde{l}_{3,{\rm min}}\simeq 2~(\widetilde{l}_{3,{\rm min}}\simeq1)$
    implies that the transition state is (is not) stabilized.
    In (a) and (b) the color bar starts from $0$,
    where $0$ is assigned to evaporation of a particle.
    (c) Theoretical stability index $S\times 10^{2}$.
    Color bar range is truncated.
    $S>0$ predicts emergence of DIC (dynamical stabilization).
    (d) Total energy.}
  \label{fig:Stability}
\end{figure}

Stability is explored at each sample point
on the plane $(\delta l_{1}/\sigma, \delta l_{2}/\sigma)$ by computing
\begin{equation}
  \widetilde{l}_{3,{\rm min}}(T) = \min_{t\in [0,T]} \dfrac{l_{3}(t)}{\ls}.
  \label{eq:l3min}
\end{equation}
We set $T=10^{2}$ and $T=10^{3}$ in Figs.~\ref{fig:Stability}(a) and (b) respectively.
The theoretical prediction by the index $S$ is shown in Fig.~\ref{fig:Stability}(c).
Numerical observation is almost in good agreement with the theoretical prediction.
The in-phase (antiphase) mode axis runs from the left-lower (right-lower) corner
to the right-upper (left-upper) corner.
Excitation of the in-phase mode stabilizes the straight conformation
if the excitation is sufficiently strong,
as we have observed in previous works \cite{yamaguchi-etal-22,yamaguchi-23}.

Slight discrepancies between numerical observation and the theoretical prediction
observed along the in-phase mode axis in particular
is again explained by asymmetry of the Lennard-Jones potential
in the contracting and extending directions.
Indeed, the total energy is asymmetric [see Fig.~\ref{fig:Stability}(d)].
$E_{\rm normal}$ is underestimated in the contracting direction,
while it is overestimated in the extending direction.
The theoretical underestimation (overestimation) induces
that numerical simulations gain (lose) stabilty around the boundary of stability.

\begin{figure}
  \centering
    \includegraphics[width=8.5cm]{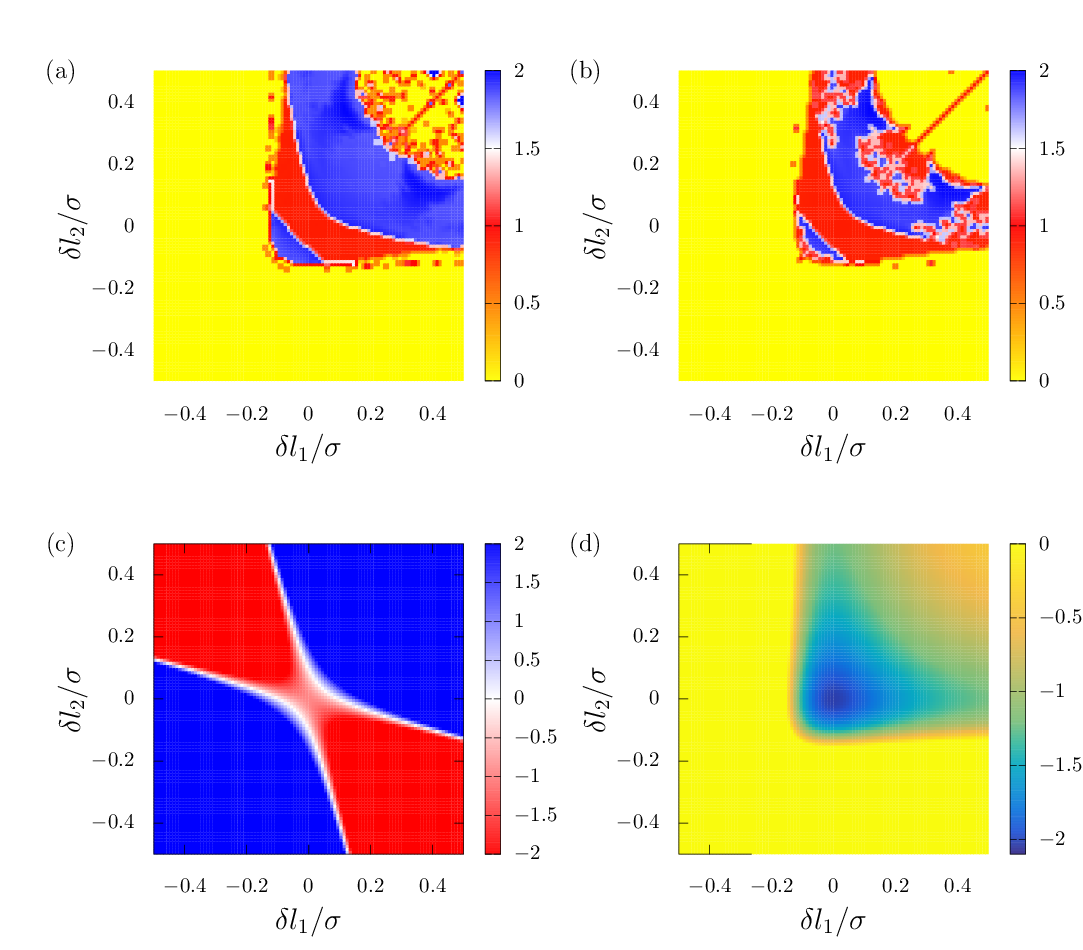}
  \caption{Same as Fig.~\ref{fig:Stability}
    but in a wider region of $(\delta l_{1}/\sigma,\delta l_{2}/\sigma)$.
    In panels (a) and (b)
    evapolation occurs at points where the color bar is $0$ (yellow).
    Color bar range is truncated in all panels.
  }
  \label{fig:Stability_wide}
\end{figure}

The same analysis is performed in Fig.~\ref{fig:Stability_wide}
for a wider range of $\delta l_{1}$ and $\delta l_{2}$.
A larger $|\delta l_{i}|$ gives a large value of the total energy
and induces frequent evaporation.
Nevertheless, DIC emerges in a upper-right region in particular.
Validity of the expansion, Eq.~\eqref{eq:expansion-l}, is no longer guaranteed in such a wider range,
but stabilization of the transition state is captured in a short time regime
except for the evaporation.

\subsection{Test of hypothesis}
\label{sec:hypothesis}

We close this section by performing a numerical test of the hypothesis:
The ratio $\nu_{\rm I}:\nu_{\rm II}$ is constant of time.
The initial condition is $(\delta l_{1},\delta l_{2})=(-0.07, -0.08)$,
which gives stability of the transition state only in a short term
(see Fig.~\ref{fig:Stability}).

The expressions of $E_{\rm I}$ and $E_{\rm II}$
[see Eqs.~\eqref{eq:EI-full} and \eqref{eq:EII-full}]
are derived by the harmonic approximation,
but this approximation is not excellent
if $|\delta l_{1}|$ and $|\delta l_{2}|$ are large (see Fig.~\ref{fig:LennardJones}).
We then compute $\nu_{\rm I}$ and $\nu_{\rm II}$
as time averages of relative amplitudes $[l_{1}^{(1)}+l_{2}^{(1)}]^{2}$
and $[l_{1}^{(1)}-l_{2}^{(1)}]^{2}$ respectively.
The averages are taken in each time window of length $5$.
For instance, the $n$th point of $\nu_{\rm I}$, denoted by $\nu_{\rm I}(n)$, is defined by
\begin{equation}
  \nu_{\rm I}(n) = \dfrac{1}{5} \int_{5n}^{5(n+1)}
  \left[ l_{1}(t) + l_{2}(t) - 2\ls \right]^{2} dt.
\end{equation}

\begin{figure}
  \centering
    \includegraphics[width=8.5cm]{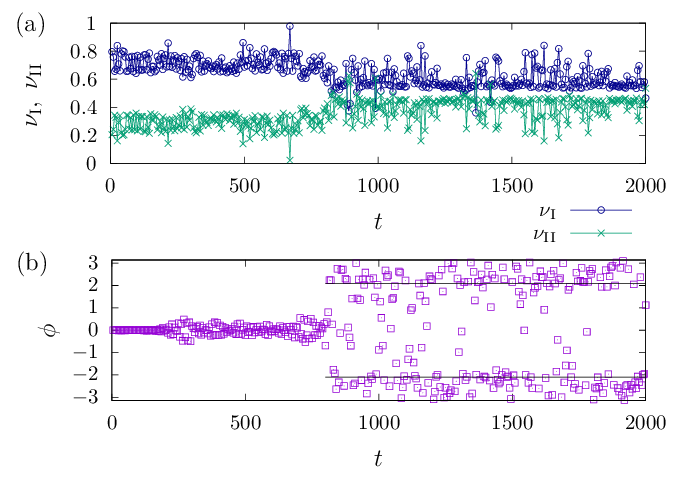}
  \caption{Hypothesis test.
    The initial condition is $(\delta l_{1},\delta l_{2})=(-0.07, -0.08)$.
    (a) Averaged temporal evolution of $\nu_{\rm I}$ (blue circles)
    and $\nu_{\rm II}$ (green crosses).
    (b) Temporal evolution of $\phi$.
    Two horizontal black lines mark $\pm 2\pi/3$ corresponding to the equilateral triangles.}
  \label{fig:Hypothesis}
\end{figure}

Temporal evolution of $\nu_{\rm I}$ and $\nu_{\rm II}$
is reported in Fig.~\ref{fig:Hypothesis} with temporal evolution of $\phi$.
They are almost constant when $\phi\simeq 0$, and go away after $t\simeq 800$
at which the system moves away from the transition state.
We hence conclude that the hypothesis is valid
while stabilization of the transition state is realized.

\section{Reaction rate}
\label{sec:transition-rate}

The transition state can be stabilized by the fast motion of $l_{1}$ and $l_{2}$.
The next question is whether the stabilization changes the reaction rate statistically.
To answer this question, we compute the reaction rate by TST and MD simulations.
The idea of TST is as follows.
We introduce a dividing surface which divides the reactant and the product in the phase space.
Then, we compute the ratio of phase volume which pass the dividing surface par unit time
to the reactant phase volume.

A more precise explanation is as follows.
Let the three bodies be A, B, and C.
In our setting, the reactant is for instance the triangle ABC,
and the product is the triangle ACB.
Let us denote the Hamiltonian of the system by $H(\by,\bp)$.
The reaction rate $k(E)$ for total energy $E$ is expressed by
\begin{equation}
  k(E) =
  \dfrac{ \displaystyle{ 3 \int_{\dot{\phi}>0} \delta(E-H(\by,\bp))\delta(\phi) \dot{\phi} d\by d\bp }}
  {\displaystyle{ \dfrac{1}{2} \int \delta(E-H(\by,\bp)) d\by d\bp }}.
\end{equation}
For the reaction route which passes the transition state $\bys=(\ls,\ls,0)$,
we may set the dividing surface as $\phi=0$,
and the numerator represents the flux passing the dividing surface in the unit time
from one side to the other.
The factor $3$ counts the three routes of isomerization
associated with the three transition states.
The denominator is the half of the phase space volume for total energy $E$,
which corresponds to the phase space volume of the reactant by symmetry.
Concrete expressions of $H(\by,\bp)$, momenta $\bp=(p^{1},p^{2},p^{3})$, and $\dot{\phi}$
are reported in Appendix \ref{sec:HamStyle}.
Recall that the total angular momentum is set as zero.

To compute $k(E)$ numerically, we truncate the phase space as
$l_{1},l_{2}\in [0.8,3], \phi=[0,2\pi), p^{1},p^{2},p^{3}\in [-P,P]$.
We examine $P=2$ and $P=3$ to confirm that $P=3$ is sufficiently large.
On each truncated axis $100$ points are equally distributed,
and the value of $H(\by,\bp)$ and the weight $\dot{\phi}$ are computed
totally at $10^{12}$ points.
Finally, we make a histogram with respect to $E$ by dividing the interval $E\in[-3,-1]$
into $100$ bins, where the representative point of the $i$th bin
is $E_{i}=-3+2(i+0.5)/100~ (i=0,\cdots,99)$.
We neglected $E>-1$ since a particle may evaporate.

The reaction rate $k(E)$ is also computed by performing direct MD simulations.
The initial condition is the one introduced in Sec.~\ref{sec:initial-condition}
with the constraint $\delta l_{2}=-\delta l_{1}$ to make the transition state unstable.
The total energy $E$ depends on $\delta l_{1}$.
The reaction rate $k(E)$ is obtained as $k(E)=N(E,T)/T$,
where $N(E,T)$ is the number of isomerization during the time interval $t\in [0,T]$,
and we set the computing time $T$ as $T=10^{4}$ and $T=10^{5}$.

\begin{figure}
  \centering
    \includegraphics[width=8.5cm]{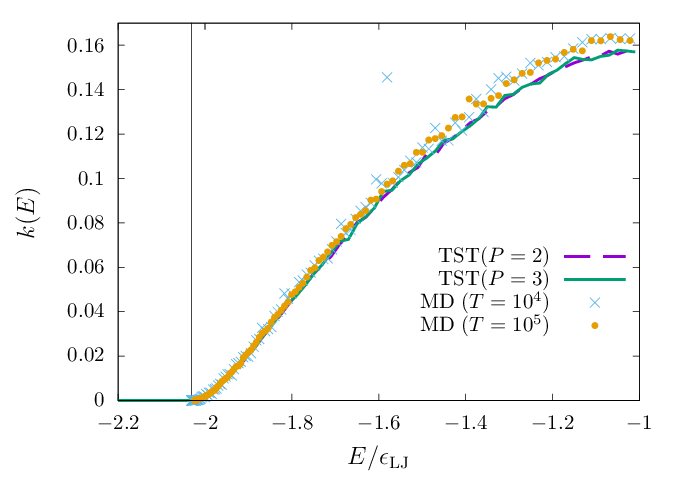}
  \caption{Reaction rate $k(E)$ as a function of $E/\epsilon_{\rm LJ}$,
    where $E$ is the total energy.
    The lines represent the transition state theory
    with $P=2$ (purple broken) and $P=3$ (green solid).
    Points are from direct molecular dynamics simulations
    with $T=10^{4}$ (blue crosses) and $T=10^{5}$ (orange circles).
    The vertical line marks the total potential energy at the transition state,
    $V(\bys)/\epsilonLJ\simeq -2.03$.}
  \label{fig:TST}
\end{figure}

The reaction rate $k(E)$ is reported in Fig.~\ref{fig:TST}.
First, we confirm that $P=3$ is sufficiently large for obtaining the theoretical precision,
since the line with $P=3$ collapses on the line with $P=2$.
Second, by a similar reason, $T=10^{5}$ is sufficiently large
for obtaining the reaction rate by MD simulations.
Therefore, we conclude that the theoretical prediction slightly underestimates
the dynamical result for $-1.6\lesssim E<-1$.
However, the theoretical prediction is not far from MD,
and the dynamical stabilization does not strongly affect the reaction rate
in the three-body Lennard-Jones system
irrespective of emergence of dynamical (de)stabilization.

\section{Summary}
\label{sec:summary}

Dynamically induced conformation (DIC) is a stabilization mechanism
of an unstable conformation with respect to a bare potential energy landscape,
and it is an analogy of the Kapitza pendulum in autonomous systems.
DIC has been previously studied in chainlike bead-spring models
with an artificially induced weak bending potential function \cite{yamaguchi-etal-22,yamaguchi-23}.
This article demonstrated that in the three-body Lennard-Jones potential
the transition state (a straight conformation) is also dynamically stabilized,
although the system is neither a chainlike system nor introducing any artificial potential.
DIC in the three-body Lennard-Jones system is based on the fact
that the potential has a steep well and a gradual tail,
and we may expect emergence of DIC in a similar system,
while a system with a harmonic pairwise potential is out of the range of the present theory.

We can derive an effective potential,
which is the sum of the bare potential and the dynamical effect
in the direction of conformation change.
The steep well provides a large prefactor in the dynamically induced term,
and it becomes comparable with instability strength of the bare potential.
However, the dynamical stabilization depends on the exited mode:
the in-phase mode contributes to stabilize the transition state,
while the antiphase mode contributes to enhance the instability,
as shown in the previous studies \cite{yamaguchi-etal-22,yamaguchi-23}.

To answer whether the dynamical (de)stabilization has a statistical impact,
we compared the reaction rate computed from the transition state theory
and from direct MD simulations.
The impact of dynamical (de)stabilization is not remarkable.
A possible explanation is coexistence of stabilization and destabilization
depending on the normal mode energy, which may be thermalized
when the system visits a well.
Nevertheless, it might be interesting to study whether
the small discrepancy between the theory and MD is physically meaningful.

Dynamics of the three-body Lennard-Jones system is considered
as the isolated system in this article.
It is worth studying DIC with thermal noises,
since they are not avoidable in a real system.
The essence of DIC is separation of scales in time and space,
and hence we may expect emergence of DIC even under thermal noises.
Another extension of the present theory is a control of molecules
by applying an external force and by using the effective local minimum around the transition state.
Such a study may give a hint to solve the excited mode dependence
of isomer population \cite{dian-longarte-zweier-02,dian-longarte-winter-zwier-04}.

\acknowledgements
The author thanks Y. Shimizu for a private communication
which informed me existence of a stabilized transition state.
This work is inspired by this communication.
The author acknowledges the support of JSPS KAKENHI Grant No. JP21K03402.

\appendix

\section{The matrix $\mB$ and the inverse matrix}
\label{sec:matrixB}
In this section, we use the symbol $\by=(l_{1},l_{2},\phi,\phi_{\rm L})$
for the internal coordinates.
The matrix $\mB(\by)$ is of $4\times 4$ accordingly.
We divide the matrix $\mB(\by)$ into $2\times 2$ matrices as
\begin{equation}
  \mB(\by) =
  \begin{pmatrix}
    \mB_{ll}(\by) & \mB_{l\phi}(\by) \\
    \mB_{\phi l}(\by) & \mB_{\phi\phi}(\by) \\
  \end{pmatrix},
\end{equation}
where
\begin{equation}
  \begin{split}
    & \mB_{ll}(\by) = \dfrac{m}{3}
      \begin{pmatrix}
        2 & \cos\phi \\
        \cos\phi & 2 \\
      \end{pmatrix},
    \\
    & \mB_{l\phi}(\by) = \dfrac{m}{3}
      \begin{pmatrix}
        - \frac{1}{2} l_{2} \sin\phi & - l_{2} \sin\phi \\
        - \frac{1}{2} l_{1} \sin\phi & l_{1} \sin\phi \\
      \end{pmatrix},
      \\
    & \mB_{\phi l}(\by) = \mB_{l\phi}(\by)^{\rm T},
    \\
    & \mB_{\phi\phi}(\by) = \dfrac{m}{3}
      \begin{pmatrix}
        \frac{1}{2}(l_{1}^{2}+l_{2}^{2}-l_{1}l_{2}\cos\phi) & l_{2}^{2}-l_{1}^{2} \\
        l_{2}^{2}-l_{1}^{2} & 2(l_{1}^{2}+l_{2}^{2}+l_{1}l_{2}\cos\phi) \\
      \end{pmatrix}.
  \end{split}
\end{equation}
The inverse matrix of $\mB(\by)$ is
\begin{equation}
  [\mB(\by)]^{-1} =
  \begin{pmatrix}
    \wt{\mB}_{ll} & \wt{\mB}_{l\phi} \\
    \wt{\mB}_{\phi l} & \wt{\mB}_{\phi\phi} \\
  \end{pmatrix},
\end{equation}
where
\begin{equation}
  \begin{split}
    & \wt{\mB}_{ll}(\by) = \dfrac{1}{m}
      \begin{pmatrix}
        2 & - \cos\phi \\
        -\cos\phi & 2 \\
      \end{pmatrix},
    \\
    & \wt{\mB}_{l\phi}(\by) = \dfrac{1}{m}
      \begin{pmatrix}
        \frac{1}{l_{2}} \sin\phi & \frac{1}{2l_{2}} \sin\phi \\
        \frac{1}{l_{1}} \sin\phi & - \frac{1}{2l_{1}} \sin\phi \\
      \end{pmatrix},
    \\
    & \wt{\mB}_{\phi l}(\by) = \wt{\mB}_{l\phi}(\by)^{\rm T},
    \\
    & \wt{\mB}_{\phi\phi}(\by) = \dfrac{1}{m}
      \begin{pmatrix}
        \frac{2}{l_{1}^{2}} + \frac{2}{l_{2}^{2}} + \frac{2}{l_{1}l_{2}} \cos\phi
        & \frac{1}{l_{2}^{2}} - \frac{1}{l_{1}^{2}} \\
        \frac{1}{l_{2}^{2}} - \frac{1}{l_{1}^{2}}
        & \frac{1}{2l_{1}^{2}} + \frac{1}{2l_{2}^{2}} - \frac{1}{2l_{1}l_{2}} \cos\phi \\
      \end{pmatrix}.
  \end{split}
  \label{eq:Binv}
\end{equation}

\section{Euler-Lagrange equations}
\label{sec:EL-eqs}

We use the $4\times 4$ matrix $\mB(\by)$ introduced in Appendix \ref{sec:matrixB}.
We expand the Euler-Lagrange equations
\begin{equation}
  \label{eq:EL}
  \sum_{\beta=1}^{4} B^{\alpha\beta} \ddot{y}_{\beta}
  + \sum_{\beta,\gamma=1}^{4} D^{\alpha\beta\gamma} \dot{y}_{\beta} \dot{y}_{\gamma}
  + \dfracp{V}{y_{\alpha}} = 0,
  \quad
  \alpha=1,2,3,4,
\end{equation}
associated with the Lagrangian \eqref{eq:Lagrangian}
into a power series of the small parameter $\epsilon$. Here,
\begin{equation}
  D^{\alpha\beta\gamma}(\by)
  = \dfracp{B^{\alpha\beta}}{y_{\gamma}}(\by) - \dfrac{1}{2} \dfracp{B^{\beta\gamma}}{y_{\alpha}}(\by),
  \quad
  \alpha,\beta,\gamma=1,2,3,4.
\end{equation}

\subsection{Expansion of the equations}
Substituting Eqs.~\eqref{eq:expansion-t}, \eqref{eq:expansion-y},
and \eqref{eq:expansion-gradV} into Eq.~\eqref{eq:EL},
we have the equations of motion in each order of $\epsilon$:
\begin{subequations}
  \begin{align}
    \label{eq:EOM-0}
    O(\epsilon^{0}): ~
    & \left( \dfracp{V}{y_{\alpha}}(\by) \right)^{(0)} = 0,
    \\
    \label{eq:EOM-1}
    O(\epsilon^{1}): ~
    & \sum_{\beta=1}^{4} B^{\alpha\beta}(\by^{(0)}) \dfracp{{}^{2}y_{\beta}^{(1)}}{t_{0}^{2}}
      + \left( \dfracp{V}{y_{\alpha}}(\by) \right)^{(1)} = 0,
    \\
    \label{eq:EOM-2}
    O(\epsilon^{2}): ~
    & \sum_{\beta=1}^{4} B^{\alpha\beta}(\by^{(0)}) \left( \ddot{y}_{\beta} \right)^{(2)}
      + \sum_{\beta,\gamma=1}^{4} D^{\alpha\beta\gamma}(\by^{(0)})
      \left( \dot{y}_{\beta} \right)^{(1)}
      \left( \dot{y}_{\gamma} \right)^{(1)} \nonumber \\
    & + \left( \dfracp{V}{y_{\alpha}}(\by) \right)^{(2)}
      + \sum_{\beta,\gamma=1}^{4}
      \dfracp{B^{\alpha\beta}}{y_{\gamma}}(\by^{(0)})
      \dfracp{{}^{2}y_{\beta}^{(1)}}{t_{0}^{2}} y_{\gamma}^{(1)} = 0,
  \end{align}
\end{subequations}
where
\begin{equation}
  (\dot{y}_{\beta})^{(1)} = \dfracd{y_{\beta}^{(0)}}{t_{1}} + \dfracp{y_{\beta}^{(1)}}{t_{0}},
  \quad
  (\ddot{y}_{\beta})^{(2)} = \dfracd{{}^{2}y_{\beta}^{(0)}}{t_{1}^{2}} + 2 \dfracpp{y_{\beta}^{(1)}}{t_{0}}{t_{1}}.
\end{equation}
The equations in $O(\epsilon^{0})$ are satisfied by Eq.~\eqref{eq:expansion-gradV}.

\subsection{Equation in $O(\epsilon)$}
\label{sec:spring-motion}

The equations of motion in $O(\epsilon)$, \eqref{eq:EOM-1}, are linear
and are rewritten as
\begin{equation}
  \label{eq:EOM-1-appendix}
  \mB^{(0)} \dfracp{{}^{2}\by^{(1)}}{t_{0}^{2}} + \mK \by^{(1)} = 0,
\end{equation}
where
\begin{widetext}
  \begin{equation}
    \mB^{(0)} = \mB(\by^{(0)})
    = \dfrac{m}{3}
    \begin{pmatrix}
      2 & \cos\phi^{(0)} & - \frac{1}{2}\ls\sin\phi^{(0)} & -\ls\sin\phi^{(0)} \\
      \cos\phi^{(0)} & 2 & - \frac{1}{2}\ls\sin\phi^{(0)} & \ls\sin\phi^{(0)} \\
      - \frac{1}{2}\ls\sin\phi^{(0)} & - \frac{1}{2}\ls\sin\phi^{(0)} & \frac{1}{2}(\ls)^{2}(2-\cos\phi^{(0)}) & 0 \\
      -\ls\sin\phi^{(0)} & \ls\sin\phi^{(0)} & 0 & 2(\ls)^{2}(2+\cos\phi^{(0)}) \\
    \end{pmatrix},
  \end{equation}
\end{widetext}
and
\begin{equation}
  \mK = \ULJ''(\ls) 
  \begin{pmatrix}
    \mE & \mO \\
    \mO & \mO \\
  \end{pmatrix}.
\end{equation}
Here $\mE$ represents the unit matrix and $\mO$ represents the zero matrix.
Substituting $\by^{(1)}=\bv\cos(\sqrt{\lambda}t_{0})$ into Eq.~\eqref{eq:EOM-1-appendix},
we consider the eigenvalue problem
\begin{equation}
  (\mB^{(0)} \lambda - \mK) \bv = \bzero,
\end{equation}
which induces $\det(\mB^{(0)} \lambda - \mK)=0$ to determine $\lambda$.
We have four sets of $\lambda_{i}$ and $\bv_{i}$ $(i=1,\cdots,4)$.
Arranging $\lambda_{i}$ into a diagonal matrix
\begin{equation}
  \mLambda = \dfrac{\ULJ''(\ls)}{m}
  {\rm diag}\left( 2-\cos\phi^{(0)}, 2+\cos\phi^{(0)}, 0, 0 \right)
\end{equation}
and $\bv_{i}$ into a matrix
\begin{equation}
  \mP =
  \begin{pmatrix}
    \bv_{1}, \cdots , \bv_{4}
  \end{pmatrix}
  =
  \begin{pmatrix}
    1 & -1 & 0 & 0 \\
    1 & 1 & 0 & 0 \\
    \frac{2}{\ls} \frac{\sin\phi^{(0)}}{2-\cos\phi^{(0)}} & 0 & 1 & 0 \\
    0 & - \frac{1}{\ls} \frac{\sin\phi^{(0)}}{2+\cos\phi^{(0)}} & 0 & 1 \\
  \end{pmatrix},
\end{equation}
we have the equality
\begin{equation}
  \mB^{(0)} \mP \mLambda - \mK \mP = \mO.
\end{equation}
We can also show that $\mP^{\rm T}\mB^{(0)}\mP = \mLambda_{B}$ is diagonal, where
\begin{widetext}
\begin{equation}
  \mLambda_{B} =
  \dfrac{m}{3} {\rm diag}\left(
    \dfrac{6}{2-\cos\phi^{(0)}},~
    \dfrac{6}{2+\cos\phi^{(0)}},~
    \frac{1}{2}(\ls)^{2}(2-\cos\phi^{(0)}),~
    2(\ls)^{2}(2+\cos\phi^{(0)})
  \right).
\end{equation}
\end{widetext}
Therefore, the matrix $\mP^{\rm T}\mK\mP=\mP^{\rm T}\mB^{(0)}\mP\mLambda
=\mLambda_{B}\mLambda$ is also diagonal.
Performing the change of variables by
\begin{equation}
  \by^{(1)}=\mP\bbeta,
  \label{eq:change-of-variables}
\end{equation}
the linear equations \eqref{eq:EOM-1-appendix} are modified into a diagonalized form
\begin{equation}
  \dfracp{{}^{2}\bbeta}{t_{0}^{2}} = - \mLambda \bbeta,
\end{equation}
where we used the fact $\det\mLambda_{B}\neq 0$.

The first two eigenvalues of $\mLambda$ are positive since $\ULJ''(\ls)>0$,
and we have two vibrating modes: the in-phase mode (mode-I) and the antiphase mode (mode-II).
Let us compute energy for each mode.
The linear equations \eqref{eq:EOM-1-appendix} are derived from the Lagrangian
\begin{equation}
  \begin{split}
    L^{(1)}
    & = \dfrac{1}{2} \left( \dfracp{\by^{(1)}}{t_{0}} \right)^{\rm T} \mB^{(0)}
      \dfracp{\by^{(1)}}{t_{0}}
      - \dfrac{1}{2} \left(\by^{(1)}\right)^{\rm T}\mK\by^{(1)} \\
    & = \dfrac{1}{2} \left( \dfracp{\bbeta}{t_{0}} \right)^{\rm T} \mLambda_{B}
  \dfracp{\bbeta}{t_{0}}
  - \dfrac{1}{2} \bbeta^{\rm T} \mLambda_{B} \mLambda \bbeta.
  \end{split}
\end{equation}
Recalling that the amplitudes of the normal modes are of $O(\epsilon)$,
we have the normal mode energy of $O(\epsilon^{2})$.
Energy of the mode-I is hence
\begin{equation}
  \label{eq:EI-eta1}
  \dfrac{E_{\rm I}}{\epsilon^{2}}
  = \dfrac{m}{2-\cos\phi^{(0)}} \left( \dfracp{\eta_{1}}{t_{0}} \right)^{2}
  + \ULJ''(\ls) \eta_{1}^{2}
\end{equation}
and of the mode-II is
\begin{equation}
  \label{eq:EII-eta2}
  \dfrac{E_{\rm II}}{\epsilon^{2}}
  = \dfrac{m}{2+\cos\phi^{(0)}} \left( \dfracp{\eta_{2}}{t_{0}} \right)^{2}
  + \ULJ''(\ls) \eta_{2}^{2}.
\end{equation}
Coming back to the coordinate $\by^{(1)}$ by $\bbeta=\mP^{-1}\by^{(1)}$ with
\begin{equation}
  \mP^{-1} =
  \begin{pmatrix}
    1/2 & 1/2 & 0 & 0 \\
    -1/2 & 1/2 & 0 & 0 \\
    -\frac{1}{\ls} \frac{\sin\phi^{(0)}}{2-\cos\phi^{(0)}} & -\frac{1}{\ls} \frac{\sin\phi^{(0)}}{2-\cos\phi^{(0)}} & 1 & 0 \\
    -\frac{1}{2\ls} \frac{\sin\phi^{(0)}}{2+\cos\phi^{(0)}} & \frac{1}{\ls} \frac{\sin\phi^{(0)}}{2+\cos\phi^{(0)}} & 0 & 1 \\    
  \end{pmatrix},
\end{equation}
we have
\begin{equation}
  \dfrac{E_{\rm I}}{\epsilon^{2}}
  = \dfrac{m[\partial_{t_{0}}(l_{1}^{(1)}+l_{2}^{(1)})]^{2}}{4(2-\cos\phi^{(0)})}
  + \dfrac{\ULJ''(\ls)}{4} (l_{1}^{(1)}+l_{2}^{(1)})^{2}
  \label{eq:EI-full}
\end{equation}
and
\begin{equation}
  \dfrac{E_{\rm II}}{\epsilon^{2}}
  = \dfrac{m[\partial_{t_{0}}(l_{1}^{(1)}-l_{2}^{(1)})]^{2}}{4(2+\cos\phi^{(0)})}
  + \dfrac{\ULJ''(\ls)}{4} (l_{1}^{(1)}-l_{2}^{(1)})^{2}.
  \label{eq:EII-full}
\end{equation}
For the initial condition discussed in Sec.~\ref{sec:initial-condition},
the initial values of $E_{\rm I}$ and $E_{\rm II}$ result in Eq.~\eqref{eq:EI-EII}
since $\delta l_{i}=\epsilon l_{i}^{(1)}$.

\subsection{Equation in $O(\epsilon^{2})$}
\label{sec:spring-motion}
Taking the average over the fast timescale $t_{0}$, we have in $O(\epsilon^{2})$
\begin{equation}
  \begin{split}
    & \sum_{\beta=1}^{4} B^{\alpha\beta} \dfracd{{}^{2}y_{\beta}^{(0)}}{t_{1}^{2}}
      + \sum_{\beta,\gamma=1}^{4} D^{\alpha\beta\gamma} \dfracd{y_{\beta}^{(0)}}{t_{1}} \dfracd{y_{\gamma}^{(0)}}{t_{1}} \\
    & = - \ave{ \left( \dfracp{V}{y_{\alpha}} \right)^{(2)} }
      - \dfrac{1}{2} \sum_{\beta,\gamma=1}^{4} \dfracp{B^{\beta\gamma}}{y_{\alpha}}
      \ave{ y_{\beta}^{(1)} \dfracp{{}^{2}y_{\gamma}^{(1)}}{t_{0}^{2}} },
  \end{split}
  \label{eq:EOM-2-ave}
\end{equation}
where $B^{\alpha\beta}, D^{\alpha\beta\gamma}$, and $\partial B^{\beta\gamma}/\partial y_{\alpha}$
are evaluated at $\by=\by^{(0)}$.
The symbol $\ave{\cdots}$ represents the average over $t_{0}$.
The left-hand side comes from the leading-order geodesic equation
in the metric space with $\mB(\by)$ [see Eq.~\eqref{eq:EL}],
the first term of the right-hand side is the averaged bare potential force,
and the second term is the additional effective force,
which is obtained by performing the integration by parts.

From now on, we focus on the equation of motion for $\phi$,
i.e. $\alpha=3$ in Eq.~\eqref{eq:EOM-2-ave}.
Around $\by=\bys$, the averaged bare potential force is
\begin{equation}
  - \ave{ \left( \dfracp{V}{\phi} \right)^{(2)} }
  = - \dfrac{1}{\epsilon^{2}} \dfracp{V}{\phi}(\ls,\ls,\phi^{(0)}),
\end{equation}
which is of $O(\epsilon^{0})$. From \eqref{eq:EOM-1-appendix},
the second term denoted by $A$ is modified into
\begin{equation}
  A = \dfrac{1}{2} \Tr \left[
    \mP^{\rm T} \dfracp{\mB}{\phi} \mP \mLambda \ave{ \bbeta \bbeta^{\rm T}} \right],
\end{equation}
where $\Tr$ represents trace. For $\bbeta=(\eta_{1},\eta_{2},\eta_{3},\eta_{4})$ we have
\begin{equation}
  \begin{split}
    & \mLambda \ave{\bbeta\bbeta^{\rm T}} 
      = \dfrac{\ULJ''(\ls)}{m} \\
    & \times {\rm diag}\left(
      (2-\cos\phi^{(0)}) \ave{\eta_{1}^{2}},
      (2+\cos\phi^{(0)}) \ave{\eta_{2}^{2}},
      0,0
      \right).
  \end{split}
\end{equation}
This form implies that, to obtain $A$, we need only the first two diagonal elements of
$\mP^{\rm T}(\partial\mB/\partial\phi)\mP$, which is read as
\begin{equation}
  \mP^{\rm T} \dfracp{\mB}{\phi} \mP =
  \begin{pmatrix}
    \begin{pmatrix}
      \frac{-2m\sin\phi^{(0)}}{[2-\cos\phi^{(0)})]{2}} & 0 \\
      0 & \frac{2m\sin\phi^{(0)}}{[2+\cos\phi^{(0)}]^{2}} \\
    \end{pmatrix}
    & \ast \\
    \ast & \ast \\
  \end{pmatrix}.
\end{equation}
The factor $A$ is hence
\begin{equation}
  A = - \ULJ''(\ls) \left( \dfrac{\sin\phi^{(0)}}{2-\cos\phi^{(0)}} \ave{\eta_{1}^{2}}
    - \dfrac{\sin\phi^{(0)}}{2+\cos\phi^{(0)}} \ave{\eta_{2}^{2}} \right).
\end{equation}
For the normal modes described by Eqs.~\eqref{eq:EI-eta1} and \eqref{eq:EII-eta2},
the averages of amplitudes are respectively
\begin{equation}
  \ave{\eta_{1}^{2}} = \dfrac{E_{\rm I}}{2\epsilon^{2}\ULJ''(\ls)},
  \quad
  \ave{\eta_{2}^{2}} = \dfrac{E_{\rm II}}{2\epsilon^{2}\ULJ''(\ls)}.
\end{equation}
Finally, from the relations
\begin{equation}
  E_{\rm I} = E_{\rm normal}\nu_{\rm I},
  \quad
  E_{\rm II} = E_{\rm normal}\nu_{\rm II},
\end{equation}
we have
\begin{equation}
  A = - \dfrac{E_{\rm normal}}{2\epsilon^{2}} \sin\phi^{(0)} \left(
    \dfrac{\nu_{\rm I}}{2-\cos\phi^{(0)}}
    - \dfrac{\nu_{\rm II}}{2+\cos\phi^{(0)}} \right).
\end{equation}
This additional term gives the effective force \eqref{eq:Feff}.

\section{Hamiltonian formalism of the system}
\label{sec:HamStyle}

The Hamiltonian of the system is expressed by
\begin{equation}
  H(\by,\bp)
  = \dfrac{1}{2} \sum_{\alpha,\beta=1}^{4} \wt{B}_{\alpha\beta}(\by) p^{\alpha} p^{\beta} + V(\by),
  \label{eq:Hamiltonian}
\end{equation}
where $\wt{B}_{\alpha\beta}(\by)$ is the $(\alpha,\beta)$ element of the inverse matrix of $\mB(\by)$
[see Eq.~\eqref{eq:Binv}].
The transformations between the momentum vector $\bp=(p^{1},p^{2},p^{3},p^{4})$
and the velocity vector $\dot{\by}=(\dot{y}_{1},\dot{y}_{2},\dot{y}_{3},\dot{y}_{4})=
(\dot{l}_{1},\dot{l}_{2},\dot{\phi},\dot{\phi}_{L})$ are
\begin{equation}
  p^{\alpha} = \sum_{\beta=1}^{4} B^{\alpha\beta}(\by) \dot{y}_{\beta},
  \quad
  \dot{y}_{\alpha} = \sum_{\beta=1}^{4} \wt{B}_{\alpha\beta}(\by) p^{\beta}.
\end{equation}
The inverse matrix $[\mB(\by)]^{-1}$ gives the velocity $\dot{\phi}$ as
\begin{equation}
  \dot{\phi} = \dfrac{1}{m} \left[
    2p^{3} \left( \dfrac{1}{l_{1}^{2}} + \dfrac{1}{l_{2}^{2}} + \dfrac{1}{l_{1}l_{2}} \cos\phi \right)
    + \left( \dfrac{p^{1}}{l_{2}} + \dfrac{p^{2}}{l_{1}} \right) \sin\phi
  \right]
\end{equation}
for the vanishing total angular momentum, $p^{4}\equiv 0$.
Similarly, the sum in the kinetic term runs from $1$ to $3$
in the Hamiltonian \eqref{eq:Hamiltonian}.

\end{document}